\documentstyle[12pt]{article}
\newcommand{\tdm}[1]{\mbox{\boldmath$#1$}}

\newlength{\dinwidth}
\newlength{\dinmargin}
\begin{document}
\titlepage
\begin{flushright}
DTP/96/18  \\
March 1996 \\
\end{flushright}

\begin{center}
\vspace*{2cm}
{\Large \bf $F_L$ at low $x$ and low $Q^2$}
\footnote{Talk presented at the Epiphany Conference, Cracow 5-6 Jan 1996.}\\
\vspace*{1cm}
A.M. Sta\'sto \footnote{On leave 
from the Department of Physics, Jagellonian University, 30-059 Cracow, Poland.}  \\
\vspace*{1.0cm}
{\it Department of Physics, University of Durham,} \\
{\it  Durham, DH1 3LE, England}
\end{center}
\vspace*{4.5cm}
\begin{abstract}
A model for the longitudinal structure function $F_L$ in the region of 
low $x$ and low $Q^2$ is discussed. It is constructed using  
the $k_T$ factorization theorem  and a photon-gluon fusion mechanism 
suitably extrapolated to the region of low $Q^2$.
A phenomenological model for higher twist is presented, which
is based on the assumption that the contribution of quarks having
limited transverse momenta is dominated by the soft pomeron exchange
mechanism.
\end{abstract}

\newpage
%%%%%%%%%%%%%%%%%%%%%%%%%%%%%%%%%%%%%%%%%%%%%%%%%%%%%%%%%%%%%%%%%%%%%%%
\section*{Introduction}

The longitudinal structure function $F_L$ corresponds to the interaction
of longitudinally polarized photons in the one--photon--exchange approximation
of deep inelastic lepton-nucleon scattering.
The experimental determination of $F_L$ is
difficult since it usually requires  measurement of the
energy dependence of this process for fixed $x$ and $Q^2$.
 The structure function $F_L$ 
vanishes within the naive parton model 
\cite{LANDSHOFF} and also  
in the leading logarithmic approximation of QCD. In the next--to--leading order approximation the structure function $F_L$
acquires a leading twist contribution proportional to $\alpha_s(Q^2)$.
The main process contributing to $F_L$ at low $x$ is 
photon--gluon fusion,  $g\gamma \rightarrow q \bar q$.
The longitudinal structure
function $F_L$ is therefore mainly sensitive to the gluon distribution.

In practice it is the ratio $R(x,Q^2)=F_L(x,Q^2)/F_T(x,Q^2)$ 
which is determined
experimentally (where $F_T(x,Q^2)$ is transverse structure function).
Current measurements of $R(x,Q^2)$ (SLAC, NMC at CERN \cite{BADELEK3} ) 
are very scarce, especially at low $x$.
In the region of large $Q^2$ the structure function $F_L$ is fairly well
described by perturbative QCD,
 while at lower values of $Q^2$ it is not well understood
mainly because of the non--perturbative contributions. 
A better understanding of $F_L$ and $R$ in this region is important
in order to implement the radiative corrections needed for the extraction of 
the $F_2 (F_3)$
structure functions from  deep inelastic scattering measurements.
It is also crucial for the reliable extraction of the 
spin dependent structure function $g_1(x,Q^2)$ \cite{BADELEK4} 
from experimental data, since the uncertainties 
in $R$ are one of the main sources of systematic errors. 

\medskip

In this paper we present a model for the structure function 
$F_L$ 
which provides extrapolation into the region of low $Q^2$. We also
propose a model for the higher twist term.
It is based on the assumption that the contribution of quarks having
limited transverse momenta is dominated by soft pomeron exchange. We introduce
an effective coupling which can be determined from the background
contribution to the structure function $F_T$ \cite{AKMS}.
Calculations for $R(x,Q^2)$ in the
region of low $Q^2$ are also presented.
This paper is based on the results presented in \cite{BBJKAS}.
%%%%%%%%%%%%%%%%%%%%%%%%%%%%%%%%%%%%%%%%%%%%%%%%%%%%%%%%%%%%%%%%%%%%%%%%%
\medskip
\section*{$k_T$ factorization vs collinear factorization}

The off--shell approximation is closely connected with the $\tdm{k_T}$
(or high energy) factorization theorem \cite{KT,KTBL} 
which can be presented in the following way,
\begin{equation}
F_L(x,Q^2)=\int_x^1\frac{dx'}{x'}\int\frac{dk_T^2}{k_T^2}
F_L^{\gamma g}(\frac{x}{x'},k_T^2,Q^2)f(x',k_T^2) .
\label{ktfakt}
\end{equation}
 The dominant diagram
for the $F_L$ is shown on Fig.1, where $x',\tdm{k_T}$
are the longitudinal fraction of proton's momentum and the transverse
momentum of the gluon $g$ respectively.
The function $F_L^{\gamma g}(\frac{x}{x'},k_T^2,Q^2)$ can be regarded 
as the longitudinal structure function of the off--shell gluon
with virtuality $k_T^2$, and
$f(x,k_T^2)$ is the unintegrated gluon distribution
which in principle should be obtained from the BFKL equation \cite{BFKL,LIP}.
In formula (\ref{ktfakt})
there is double convolution over the variables $x'$ and $\tdm{k_T}$, and
therefore it is called the off--shell approach.
On the other hand
the well known renormalization group analysis is basically
connected with collinear factorization formula,
\begin{equation}
F(x,Q^2)=\sum_i\int_x^1\frac{dx'}{x'}C_i(x/x',\alpha_s,Q^2/\mu^2)x'h_i(x',\mu^2)\;.
\label{kolfakt}
\end{equation}
The $h_i(x',\mu^2)$ are the integrated parton distribution functions which
satisfy conventional Altarelli-Parisi equations \cite{AP}
and $C_i$ are the coefficient functions which can be calculated perturbatively.
The summation is performed over different species of partons and
$\mu^2$ is a factorization scale which usually is taken to be $Q^2$ 
in case of deep inelastic scattering.

It has been shown in \cite{KT,KTJKAM} that the leading twist part
of the $\tdm{k_T}$
factorization formula can be recast into the collinear form (\ref{kolfakt})
 where all small $x'$ $(x/x')$
effects are resummed to all orders in $\alpha_sln(x')$ $(\alpha_sln(x/x'))$
in the splitting (and coefficient) functions.
For instance the moments of the coefficient functions
$\bar C_L(\omega,\alpha_s)$
\begin{equation}
\bar C_L(\omega,Q^2,\alpha_s(Q^2))\equiv\int_0^1{dz \over z} z^{\omega}
C_L(z,Q^2,\alpha_s(Q^2))
\end{equation}
are related to the double moment function $\tilde F_L^{\gamma g}(\omega,\gamma)$
 by
\begin{equation}
\bar C_L(\omega,Q^2,\alpha_s(Q^2))=\gamma_{gg}(\bar \alpha_s/\omega)
\tilde F_L^{\gamma g}(\omega,\gamma=\gamma_{gg}(\bar \alpha_s/\omega))
\end{equation}
where $\bar\alpha_s=N_c\alpha_s/\pi$, and
\begin{eqnarray}
\bar F_L^{\gamma g}(\omega,Q^2)=
{1 \over 2\pi i}\int_{1/2-i\infty}^{1/2+i\infty}d\gamma
\tilde F_L^{\gamma g}(\omega,\gamma)
\left({Q^2 \over k^2}\right)^{\gamma}   \nonumber \\
\bar F_L^{\gamma g}(\omega,Q^2)=\int_0^1{dx \over x}x^{\omega} 
F_L^{\gamma g}(x,Q^2)\;.
\end{eqnarray}
The quantity $\gamma_{gg}(\bar \alpha_s/\omega)$ is the gluon anomalous dimension
related to the appropriate splitting function by
\begin{equation}
\gamma_{gg}(\bar \alpha_s/\omega)=\int_0^1dz z^{\omega}P_{gg}(z,\alpha_s)\;.
\label{splitting}
\end{equation} 
Knowing the expansion of $\tilde F_L^{\gamma g}$ in terms of the 
anomalous dimension and the expansion
of $\gamma_{gg}$ in powers of $\bar \alpha_s/\omega$, one can obtain the following formula
for the coefficient function $C_L(z,\alpha_s)$ (in case of $F_L$),
\begin{equation}
C_L(z,\alpha_s)=\alpha_s[C_L^{(0)}(z)+\alpha_s C_L^{(1)}(z)
+\alpha_s^2 ln({1 \over z}) C_L^{(2)}(z)
+\alpha_s^3 ln^2({1 \over z}) C_L^{(3)}(z)+ \ldots] 
\label{coefexp}
\end{equation}
where as $z \rightarrow 0$, $C_L^{(0)}(z) \rightarrow 0 \hspace{1.5cm}
C_L^{(i)}(z) \rightarrow const \neq 0, \hspace{0.5cm} i \geq 1$.

If instead of the BFKL solution in  (\ref{ktfakt}), we take
\begin{equation}
f(x,k_T^2)=\left. \frac{\partial xg(x,Q^2)}{\partial lnQ^2}\right|_{Q^2=k_T^2}
\label{uninap}
\end{equation}
where $xg(x,Q^2)$ is a solution of the Altarelli-Parisi equations
without small $x$ corrections in $P_{gg}$ 
but preserving the original phase space (no $\tdm{k_T}$ ordering),
then only the first three non-trivial corrections in small $x$
are included in the expansion for $C_L$.
This approximation seems to be reliable, but not of course at very
low values of $x$ where the BFKL
effects could play an important role. In our calculations we consider
$x>10^{-4}$.

%%%%%%%%%%%%%%%%%%%%%%%%%%%%%%%%%%%%%%%%%%%%%%%%%%%%%%%%%%%%%%%%%%%%%%%%%
\medskip
\section*{The photon - gluon fusion model for $F_L$.}
Here we show how after an expanding the integrand
in the off--shell
 formula (\ref{ktfakt}) around $k_T^2=0$
one can obtain the on--shell approximation \cite{RGR} with additional power
corrections.
If we use the $\tdm{k_T}$ factorization theorem to calculate $F_L$,
then we obtain the following form for $F_L$ at
low $x$ and large $Q^2$:
\begin{equation} 
F_L(x,Q^2)=2 {Q^4\over \pi^2} \sum _qe_q^2 I_q(x,Q^2)
\label{flsum} 
\end{equation}
where
\begin{equation} 
I_q(x,Q^2)=\int {dk_T^2\over k_T^4} \int_0^1 d\beta \int d^2\kappa'_T
\alpha_s(Q^2) \beta^2 (1-\beta)^2 {1\over 2}\left({1\over D_{1q}} - {1\over D_{2q}}
\right)
^2f(x',k_T^2) 
\label{flint}
\end{equation} 
and 
$$ D_{1q}=\kappa_T^2 + \beta (1-\beta) Q^2 + m_q^2 $$
\begin{equation}
 D_{2q}=(\tdm{\kappa_T}-\tdm{k_T})^2 + \beta (1-\beta) Q^2 + m_q^2\;.
\label{d} 
\end{equation}
The transverse momentum $\tdm{\kappa^{\prime}_T}$ and the variable
$x'$ are defined as follows: 
\begin{equation}
\tdm{\kappa^{\prime}_T} =\tdm{ \kappa_T}- (1-\beta)\tdm{k_T} 
\label{kappap}
\end{equation}
\begin{equation}
x'=x\left(1 + {\kappa^{\prime 2}_T + m_q^2 \over \beta (1-\beta)Q^2} + 
{k_T^2\over Q^2}\right) \;.
\label{yq}
\end{equation}
The integration over $x'$ is implicit in (\ref{flint}). One can make
it explicit by changing the variables from $\kappa'^2_T$ to $x'$ using
the above formula (\ref{yq}).
The integration limits in (\ref{flint}) are additionally constrained by the 
condition:
\begin{equation} 
x'<1\;.
\label{y1}
\end{equation}
The condition $x'>x$ demanded by the integration limit 
is automatically satisfied, see (\ref{yq}). 
In the on--shell approximation of the formulae
(\ref{flsum},\ref{flint})
 one expands 
 integrands of the corresponding integrals  around $k_T=0$ 
and retains the leading term.     
The integrals $I_q$ 
can then be expressed in terms of  the conventional integrated 
gluon distribution 
$g(y,Q^2)$
\begin{equation}
I_q(x,Q^2)= \pi \int_0^1 d\beta  \int d\kappa_T^{\prime 2} 
\alpha_s(Q^2) \beta^2 (1-\beta)^2 {\kappa_T^{\prime 2}\over D_q^4}
yg(y,Q^2) \;,
\label{iq}
\end{equation}
where now
\begin{equation}
y=x\left(1 + {\kappa_T^{\prime 2} + m_q^2 \over \beta (1-\beta)Q^2}\right) 
\label{y}
\end{equation}
and
\begin{equation}
D_q= \kappa_T^{ \prime 2} + \beta (1-\beta) Q^2 + m_q^2\;.
\label{dq}
\end{equation}
In order to extrapolate formula (\ref{flsum})
 (with $I_q$ given by (\ref{flint})
or (\ref{iq}) ) to the low $Q^2$ region
we have to freeze the evolution of $g(y,Q^2)$ and the argument 
of $\alpha_s(Q^2)$.  To do this we  
  substitute $Q^2+4m_q^2$ instead of $Q^2$ as the argument of
$\alpha_s$ and of $yg(y,Q^2)$.    
\par
This substitution may be justified by analyticity arguments i.e. 
we want $I_q$ to have threshold singularities for $Q^2< -4m_q^2$.  
Moreover, for heavy quarks and for moderately large values of $Q^2$,
 it is the heavy mass squared and not $Q^2$ 
which should be taken as the scale of $\alpha_s$ and of the parton 
distributions. It should be noted that after these modifications the 
structure function $F_L$ can be continued down to $Q^2=0$,
respecting the kinematical constraint  $F_L \rightarrow Q^4$ as $Q^2
\rightarrow 0$. 
The formula (\ref{iq})  can be rearranged to 
give the Altarelli-Martinelli \cite{AM} integral over $y$ but with the modified 
integration limits and additional power corrections.  
First we use $y$ and $\beta$ as the integration 
variables instead of $\kappa^{\prime 2}_T$ and $\beta$.  
From (\ref{dq}) we find
\begin{equation}
 \kappa^{\prime 2}_T=\beta(1-\beta)Q^2({y\over x}-1)-m_q^2 
\label{kappap2}
\end{equation}
\begin{equation}
D_q=\beta(1-\beta)Q^2{y\over x}\;.
\label{dq2}
\end{equation}
From eq. (\ref{kappap2}) we can deduce various integration limits since
\begin{equation}  
\beta(1-\beta)Q^2({y\over x}-1)-m_q^2>0 
\label{limit1}
\end{equation}
and since $1>\beta>0$,
\begin{equation}
{1\over 4} >\beta(1-\beta)> {m_q^2\over Q^2({y\over x}-1)}\;.
\label{limit2}
\end{equation}
From inequality (\ref{limit2}) we obtain the lower limit for integration 
over $y$
\begin{equation}
y>x(1+{4m_q^2\over Q^2})
\label{limity}
\end{equation}
and of course,
\begin{equation} 
\beta(1-\beta)>{m_q^2\over Q^2({y\over x}-1)} 
\label{betal}
\end{equation}
It is convenient to make another change of the integration 
variables: 
\begin{equation}
\beta={1\over 2} +\lambda\;.
\label{lambda}
\end{equation}  
The inequality (\ref{limit2}) gives the following limits for the variable 
$\lambda$
\begin{equation} 
-\sqrt{{1\over 4}-{m_q^2\over Q^2({y\over x}-1)}} <\lambda<      
\sqrt{{1\over 4}-{m_q^2\over Q^2({y\over x}-1)}}\;.
\label{lamdal}
\end{equation}
Finally we get the following representation for $F_L$: 
\begin{eqnarray}
F_L=2\sum_q e_q^2 (J_q^{(1)} - 2{m_q^2\over Q^2} J_q^{(2)})
\label{iq2}
\end{eqnarray} 
where
\begin{eqnarray} 
J_q^{(1)}&=&{\alpha_s \over \pi}\int_{\bar x_q}^1{dy\over y}
\left({x\over y}\right)^2
\left(1-{x\over y}\right)\sqrt{1-{4 m_q^2 x\over Q^2({y-x})}}\,yg(y,Q^2)
\label{iq3} \\
J_q^{(2)}&=&{\alpha_s \over \pi} \int_{\bar x_q}^1
{dy\over y}\left({x\over y}\right)^3
ln \left({1+\sqrt{1-{4 m_q^2 x\over Q^2(y-x)}}\over 
1-\sqrt{1-{4 m_q^2 x\over Q^2(y-x)}}}\right)yg(y,Q^2) 
\label{iq4}
\end{eqnarray}
and
\begin{equation}
\bar x_q \equiv x\left(1+\frac{4m_q^2}{Q^2}\right)\;.
\label{xq}
\end{equation}

It should be noted that expressions (\ref{iq2},\ref{iq3},\ref{iq4})
reduce to the following formula
\begin{equation}
F_L(x,Q^2)={\alpha_s(Q^2) \over \pi} \int_x^1 {dy \over y} \left({x \over y}
\right)^2\left(1-{x \over y}\right)yg(y,Q^2)
\label{artmart}
\end{equation}
 given
by \cite{AM} in the \mbox{$Q^2 \rightarrow \infty$} limit. The expression 
$\frac{m_q^2}{Q^2}J_q^{(2)}$ represents the power correction enhanced by
the logarithmic factor $lnQ^2/m_q^2$.
 The modification of the lower limit becomes obvious, if we consider
the energy--momentum conservation constraint for the subprocess
$\gamma g \rightarrow q \bar q$.
\medskip
%%%%%%%%%%%%%%%%%%%%%%%%%%%%%%%%%%%%%%%%%%%%%%%%%%%%%%%%%%%%%%%%%%%%%%%%%
\section*{A phenomenological model for the higher--twist contribution to $F_L$}

The kinematical constraint restricts the behaviour of $F_L$ in the
region of small values of $Q^2$. 
Analysing the longitudinal cross
section for deep inelastic scattering one concludes that in order to
remove the possible singularities of this observable 
at $Q^2 \rightarrow 0$,
(i.e. $\sigma_L \sim F_L/Q^2$ ), $F_L$ should
behave as $Q^4$. This implies that $\sigma_L \sim Q^2$,
so it vanishes in the real photoproduction limit as expected.
Phenomenological analysis of the quantity $R$ in the region
 of moderately large $Q^2$ ( $Q^2>1GeV^2$ ) implies that the
higher twist effects (i.e. terms that behave like $1/Q^2$ as 
$Q^2 \rightarrow \infty$) should play an important role \cite{BODEK}. 

In this paragraph we introduce a phenomenological
model for the higher twist term which satisfies the constraint $F_L \sim Q^4$
as $Q^2 \rightarrow 0$.
The idea is to treat the contributions from
low and high quark momenta in a different way. We divide the
integration over $\kappa'^2_T$ (in (\ref{flint}) for the off--shell model
(or in (\ref{iq3}),(\ref{iq4}) for the on--shell model ) into two parts --
the region
 $0<\kappa'^2_T<\kappa'^2_{0T}$, and $\kappa'^2_T>\kappa'^2_{0T}$ --
  where $\kappa'^2_{0T}$ is
an arbitrary cut--off, chosen to be of order $1GeV^2$.
The region of high momenta is treated in the usual way (using off--shell
(\ref{flsum},\ref{flint}) 
or on--shell (\ref{iq2},\ref{iq3},\ref{iq4}) formula)
 whereas for low momenta we assume
the existence of an effective coupling constant:
\begin{equation}
\alpha_s(Q^2)zg(z,Q^2)\rightarrow A\;.
\label{effconst}
\end{equation}
The constant $A$ is not a free parameter. We can estimate it from $F_2$
assuming that the non-perturbative contribution to $F_2$ in the
scaling region also comes from the region of low values of $\kappa_T'^2$
and is controlled by the same parameter.
The contribution to $F_L$ from the region
$0<\kappa'^2_T<\kappa'^2_{0T}$
is calculated using on--shell formula, 
making the substitution (\ref{effconst})
and integrating over $\kappa'^2_T$,
\begin{eqnarray}
\Delta F_L(Q^2)&=&A\frac{2}{\pi}Q^4\sum_q e_q^2\int_0^1d\beta\beta^2(1-\beta)^2
\left\{\frac{1}{6}\frac{1}{[\beta(1-\beta)Q^2+m_q^2]^2}-\right. \nonumber\\
&& -\frac{1}{6}\left.\frac{\beta(1-\beta)Q^2+m_q^2+3\kappa'^2_{0T}}
{[\beta(1-\beta)Q^2+m_q^2+\kappa'^2_{0T}]^3}\right\}\;.
\end{eqnarray}
This formula vanishes like $1/Q^2$ for large values of $Q^2$, therefore
it has the form of a higher twist term. It also obeys
the kinematical constraint for the structure function $F_L$ ,
i.e. vanishes like $Q^4$ for $Q^2\rightarrow0$.

To evaluate the constant $A$ we consider the corresponding formula for $F_T$,
\begin{eqnarray}
F_T(x,Q^2)&=&2\sum_q e_q^2\frac{Q^2}{4\pi^2}\int_0^1d\beta \int d^2{\kappa'}_T 
\alpha_s ({\kappa'}^2_T+m_0^2)\times \nonumber\\
&& \times\left\{[\beta^2+(1-\beta)^2]\left[\frac{\kappa_T^2}{D^2_{1q}}-\frac{\tdm{\kappa_T}
\cdot(\tdm{\kappa_T}-\tdm{k_T})}{D_{1q}D_{2q}}\right]+\frac{m_q^2}{D^2_{1q}}-\frac{m_q^2}
{D_{1q}D_{2q}}\right\}f(x',k_T^2) \nonumber\\
\label{offshellft}
\end{eqnarray}
where $\tdm{\kappa'_T}$, $D_{1q}$ and $D_{2q}$ are defined by eqs. (\ref{d},
\ref{kappap}).
If we rewrite the integrands to be symmetric in variables 
$\tdm{\kappa_T}$ and $\tdm{\kappa_T}-\tdm{k_T}$ we obtain,
\begin{eqnarray}
\left[\frac{\kappa_T^2}{D^2_{1q}}-\frac{\tdm{\kappa_T}
\cdot(\tdm{\kappa_T}-\tdm{k_T})}{D_{1q}D_{2q}}\right]=
\frac{1}{2}\left[\frac{\tdm{\kappa_T}}{D_{1q}}-\frac{\tdm{\kappa_T}
-\tdm{k_T}}{D_{2q}}\right]^2 \nonumber\\
\frac{m_q^2}{D^2_{1q}}-\frac{m_q^2}{D_{1q}D_{2q}}=\frac{m_q^2}{2}\left[\frac{1}{D_{1q}}-
\frac{1}{D_{2q}}\right]^2
\label{eqkap}
\end{eqnarray}
If we now expand the right hand side of the eq.(\ref{eqkap})
in powers of $k_T^2/Q^2$ ( on--shell approximation) we obtain
the following formula,
\begin{eqnarray}
\left[\frac{\tdm{\kappa_T}}{D_{1q}}-\frac{\tdm{\kappa_T}
-\tdm{k_T}}{D_{2q}}\right]^2 \simeq \left[\frac{\tdm{k_T}}{D_q}-
\frac{2\tdm{\kappa'}_T(\tdm{\kappa'}_T\cdot \tdm{k_T})}{D_q}\right]^2= \nonumber\\
=k_T^2\left(\frac{1}{D_q^2}-\frac{4cos^2\phi\kappa'^2_T}{D_q^3}+
\frac{4cos^2\phi\kappa'^4_T}{D_q^4}\right)
\end{eqnarray}
where we have dropped terms containing odd powers of $cos\phi$ which give
vanishing contribution after performing intergation over $d\phi$.
Then $F_T$ takes the form
\begin{eqnarray}
F_T(x,Q^2)&=&2\, A\sum_q e_q^2\frac{Q^2}{4\pi}\int_0^1d\beta
\int_0^{\kappa'^2_{0T}} d{\kappa'}^2_T 
\nonumber\\
&& \times\left\{\frac{1}{2}[\beta^2+(1-\beta)^2]\left(\frac{1}{D_q^2}-\frac{2\kappa'^2_T}
{D_{q}^3}+\frac{2\kappa'^4_T}{D_{q}^4}\right)+\frac{m_q^2\kappa'^2_T}{D_q^4}\right\}\;.
\end{eqnarray}
Performing the integration over $\kappa'^2_T$, substituting 
(\ref{effconst}) and taking the  limit $Q^2\rightarrow\infty$ ,
we get the following form for the formula for the structure function $F_L$,

\begin{equation}
A=\frac{2\pi{F_T^{bg}}}{\sum_qe_q^2\left\{-\frac{2}{3}ln\frac{m_q^2}{m_q^2+\kappa'^2_{0T}}+
\frac{1}{3}\frac{\kappa'^2_{0T}}{m_q^2+\kappa'^2_{0T}}\right\}}\;.
\end{equation}
The value for $F_T^{bg}$ is taken from the phenomenological model for $F_T$
\cite{AKMS} where $F_T$ is assumed to have the form,
\begin{equation}
F_2 \simeq F_T=F_T^{bg}+F_T^{BFKL}
\end{equation}
where  $F_T^{bg}$ is the background originating from non-perturbative
region and its value is fixed by the data to be around $0.4$.
Assuming $\kappa'^2_{0T}=1GeV^2$ one obtains $A \simeq 1.96$.

%%%%%%%%%%%%%%%%%%%%%%%%%%%%%%%%%%%%%%%%%%%%%%%%%%%%%%%%%%%%%%%%%%%%%%%%%%%%%%
\medskip
\section*{Numerical results}
In this section we present the numerical analysis for the structure
function $F_L$. We shall also determine the ratio $R$
and compare it with SLAC parametrisation \cite{SLACR}.
We include $u,d,s,c$ quark contributions, with quark masses $0.35, 0.35, 0.5,
1.5 \; GeV$ respectively.
We have used both GRV \cite{GRV} and MRS(A) \cite{MRSA} parton distributions.
In GRV the LO gluons and coupling constant were used.
The LO coupling constant has the following form,
\begin{equation}
{\alpha_s(Q^2) \over 2\pi}={2 \over \beta_0ln{Q^2 \over \Lambda^2}}
\label{coupl}
\end{equation}
where $\beta_0=11-2N_f/3$, $N_f$ is the number of flavours, and the value
of the constant $\Lambda$ depends on number of active flavours
\begin{equation}
\Lambda^{(3,4)}_{LO}=0.232,0.200 \; GeV\;.
\end{equation}
We include the charm quark production threshold  i.e.
if $Q^2>4m_c^2$ then $N_f=4$, otherwise $N_f=3$.

In order to extrapolate the structure function $F_L$ to low $Q^2$
we use the method of freezing the argument of the coupling and
of the gluon distributions as explained before.
For the off--shell formula we have used the prescription proposed in \cite{BLUEMLEIN}.
In order to regularise the integration over $k_T^2$ we have
introduced the cut-off $k_{0T}^2$ .
The contribution of integration from $0$ to $k_{0T}^2$ is calculated
using the approximate formula (\ref{iq2},\ref{iq3},\ref{iq4}) with the substitution
$yg(y,k_{0T}^2)$. The magnitude of the structure function $F_L$
turns out to depend rather weakly on the chosen cut-off $k_{0T}^2$.
The dominant contribution to $F_L$ is given by the photon-gluon fusion
subprocess $\gamma g \rightarrow q\bar{q}$
 (so the gluon distribution is the most important at low $x$) but the
contributions from quarks have also been included with the modification
$x\rightarrow x(1+4m_q^2/Q^2)$ for the lower limit of the integration over $y$.
\begin{equation}
\Delta F_L(x,Q^2)=\sum_i e_i^2 \frac{\alpha_s(Q^2+4m_i^2)}{\pi}
\frac{4}{3}\int_{\bar x_q}^1\frac{dy}{y}
(\frac{x}{y})^2 [q_i(y,Q^2+4m_i^2)+\bar q_i(y,Q^2+4m_i^2)]
\end{equation}
where $\bar x_q$ is given by (\ref{xq}).
In Fig.2. we compare our results for massive and massless
quarks. As can be observed the quark masses play an important role
in the region of low $Q^2$. But even at  moderate values of $Q^2$
the difference is visible. This is because the on--shell model
(\ref{iq2},\ref{iq3},\ref{iq4})  with
masses contains the power correction term multiplied by the logarithmic factor, 
$\sim m^2/Q^2ln(Q^2/m^2)$. Note also that the difference
between the on--shell and off--shell calculations is very small.

In Fig.3 we present $F_L$ as a function of $x$,
with and without higher twist corrections.
We have chosen the cut-off parameter $\kappa'^2_{0T}$  to be
$1.0GeV^2$, but there is no
significant dependence on this quantity
(within the range of $0.8<\kappa'^2_{0T}<2.0GeV^2$).
Higher twist gives higher values for $F_L$, especially in the region
where $x$ is not very small ($x>0.001$). On the other hand it gives
lower values for small $x$ than the model without higher twist.
This is due to the fact that the contribution coming 
from low values of the quark momenta is now described by the
soft pomeron exchange.
The difference between these calculations decreases as $Q^2$ increases.

In Fig.4 we present the results for the ratio $R(x,Q^2)=F_L(x,Q^2)/
(F_2(x,Q^2)-F_L(x,Q^2))$ together with SLAC \cite{SLACR}
parametrisation.
For both parton sets (GRV and MRS(A)) the structure  function $F_2$
has been calculated from the VMD model \cite{BBJK}.
We also observe the same effect as for the longitudinal structure function:
the model with higher twist predicts bigger values for the ratio $R$
at high $x$.

For intermediate values of $Q^2$ this model coincides with
SLAC parametrisation, whereas for lower values of $Q^2$ it vanishes quickly
as $Q^2 \rightarrow 0$.
It should be noted that the difference between the parametrisations
in the low $Q^2$ region are not very large, they both decrease very
fast with decreasing $Q^2$. 
This suggests that the ratio $R$ and the structure function $F_L$
 are not very sensitive
to the differences in parton parametrisations in this region. Because the
SLAC model is not applicable in the low $Q^2$ region, using it 
as an estimate of the ratio $R$ may cause significant errors.
%%%%%%%%%%%%%%%%%%%%%%%%%%%%%%%%%%%%%%%%%%%%%%%%%%%%%%%%%%%%%%%%%%%%%%%%5
\medskip
\section*{Summary}
We have analysed the proton structure function $F_L$
in the low $Q^2$ and low $x$ region.
We have studied the off--shell formula which originates from the high 
energy factorization theorem and we have proved that in the limit when
$k_T^2 \rightarrow 0$ it reduces to the on--shell approximation.
The numerical differences between these two approaches are not very 
significant.
We have also studied the structure function $F_L$ in the very low $Q^2$ region, where
we have proposed the method  of continuing the formulae
for the structure functions down to $Q^2=0$. The non-zero values
of quark masses have strong consequences in the structure function
behaviour. Both the off--shell and on--shell formulae contain
power corrections connected with quark masses (i.e. terms like $m^2/Q^2
ln(Q^2/m^2)$ ). We have also proposed a model for the higher twist term
which is based on the assumption that the contribution of quarks having
limited transverse momenta is dominated by soft pomeron exchange.
 This results
in change of the $x$ dependence of the structure function $F_L$. 
We have also performed calculations for $R$. The model
for the ratio $R$ enables us to estimate this quantity in the low $Q^2$ region.
It is particularly important because the uncertainty in the ratio $R$ is the main
source of systematic errors when extracting other structure functions
from experimental data.
%%%%%%%%%%%%%%%%%%%%%%%%%%%%%%%%%%%%%%%%%%%%%%%%%%%%%%%%%%%%%%%%%%%%%%%%%%
\section*{Acknowledgments}
I am very grateful to Jan Kwieci\'nski and Barbara Bade\l{}ek
for collaboration and for many fruitful discussions. 
I thank Alan Martin for critically reading the manuscript
and useful comments.
I would like to thank the
Physics Department and Grey College of the University of Durham
for their warm hospitality.

\medskip
%%%%%%%%%%%%%%%%%%%%%%%%%%%%%%%%%%%%%%%%%%%%%%%%%%%%%%%%%%%%%%%%%%%%%%%%%%%

%%%%%%%%%%%%%%%%%%%%%%%%%%%%%%%%%%%%%%%%%%%%%%%%%%%%%%%%%%%%%%%%%%%%%%%%%%%%
\section*{Figure captions}

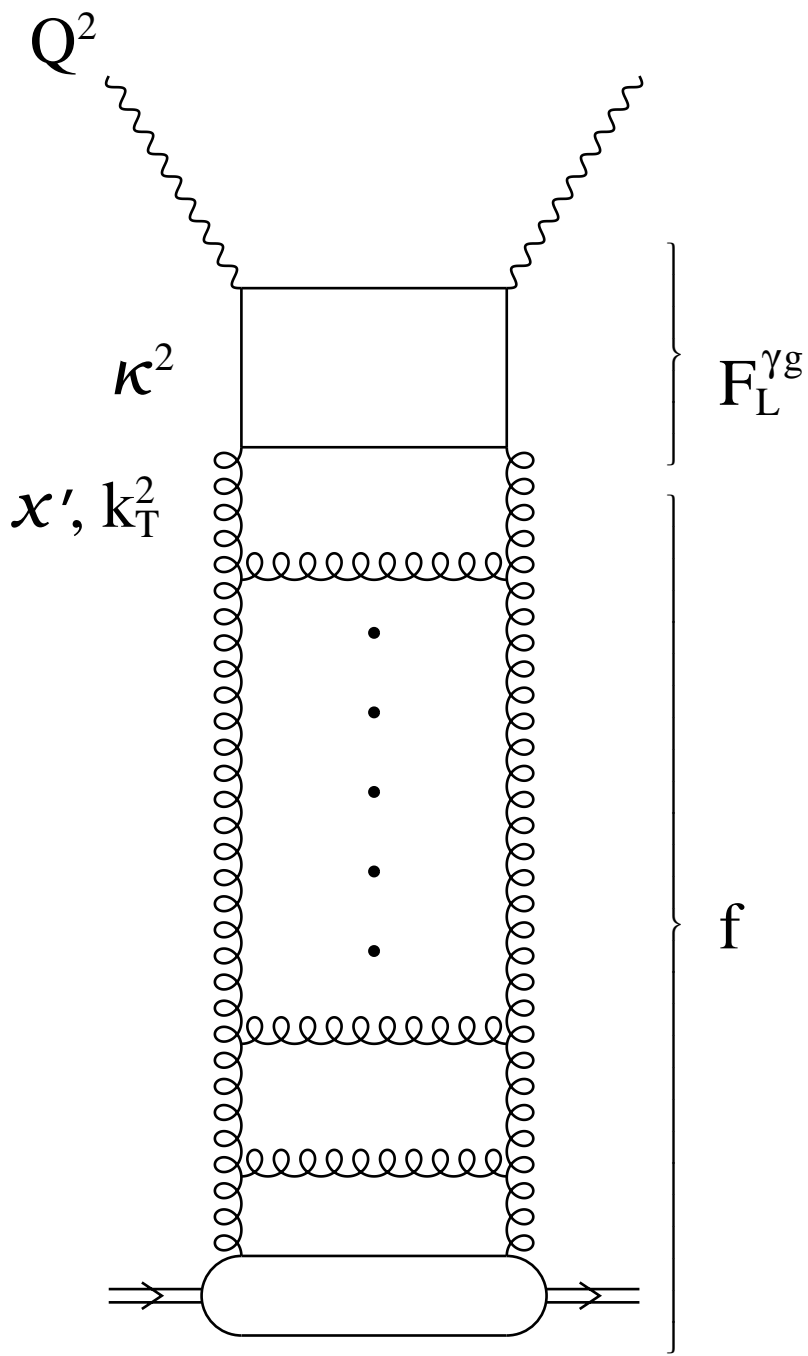
 {\bf Fig.1} The diagramatic representation of the $k_T$
factorization formula for $F_L$. \\
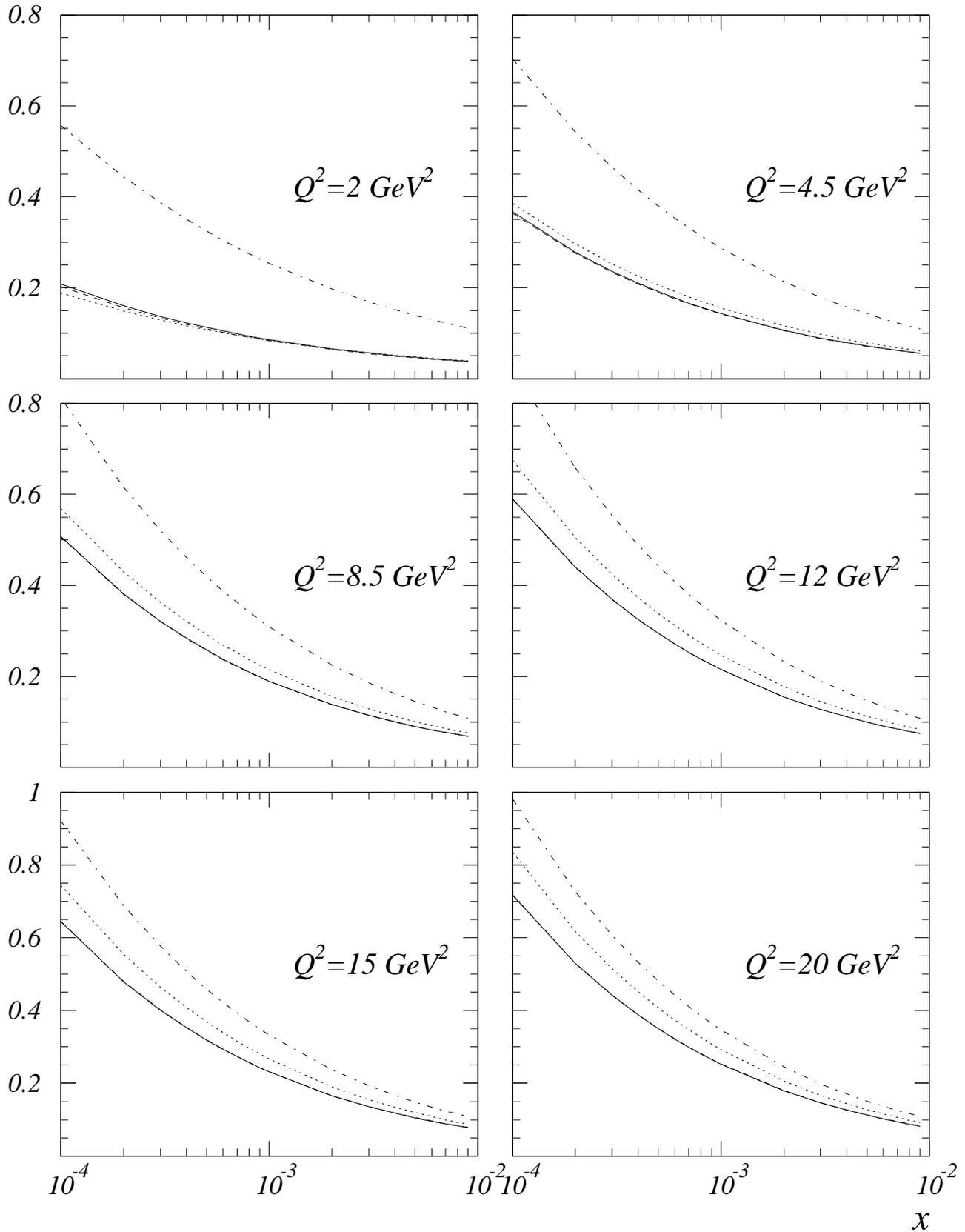
{\bf Fig.2} $F_L$ plotted as a function of $x$
 for different values of $Q^2$.
We use GRV partons \cite{GRV}.
 Solid line: off--shell approximation
with $k_{0T}^2=0.23GeV^2$, dashed line: off--shell approximation
with $k_{0T}^2=1.0GeV^2$, dotted line: on--shell approximation,
 dashed-dotted line on--shell approximation with massless
quarks {\it u,d,s}.\\
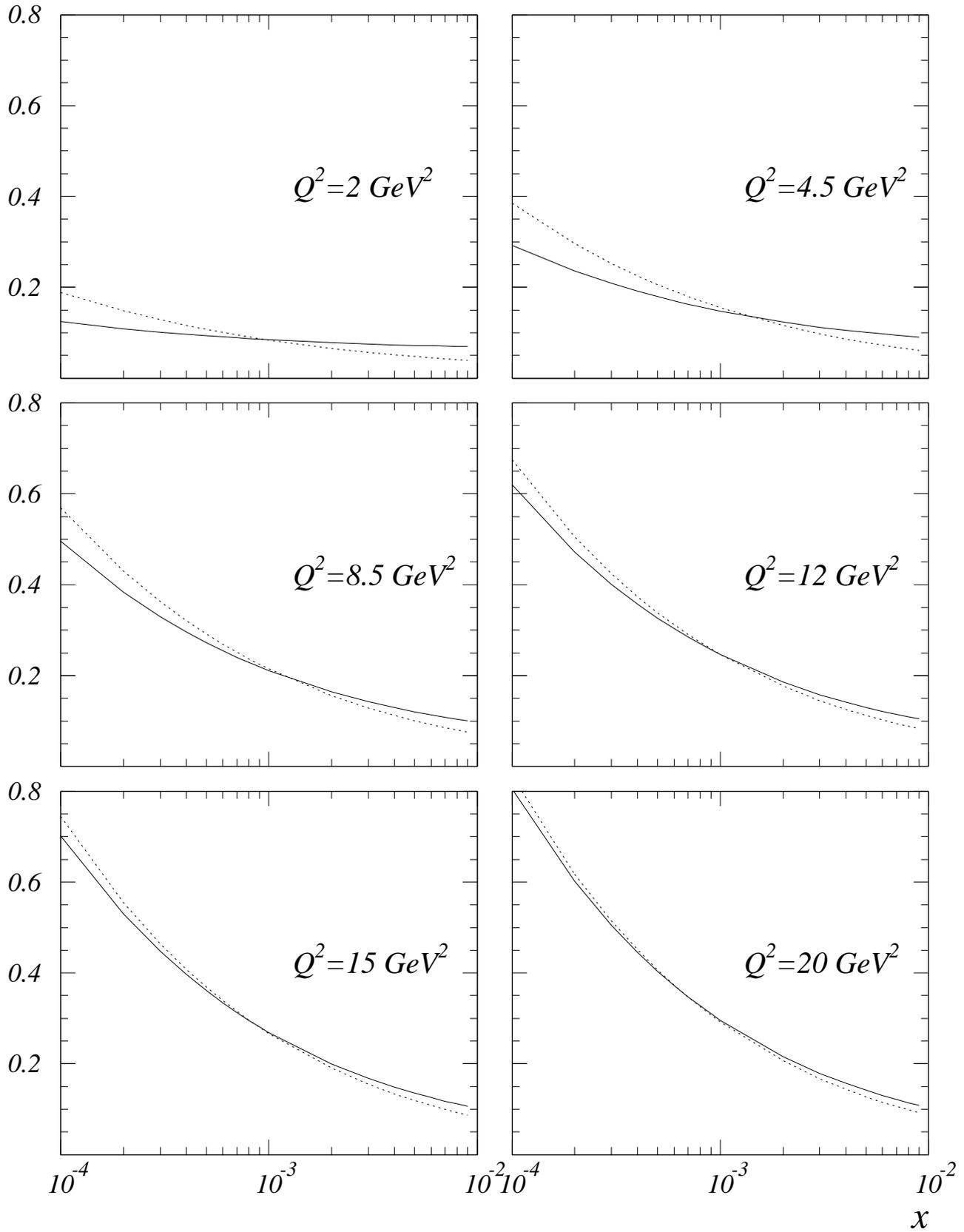
{\bf Fig.3} $F_L$ as a function of $x$ for different values of $Q^2$
calculated in the on--shell approximation:
comparison of higher twist effect, partons distributions are obtained using
the GRV parametrisation.
 Solid and dotted lines  correspond to the results without 
 and with the higher twist term respectively.\\
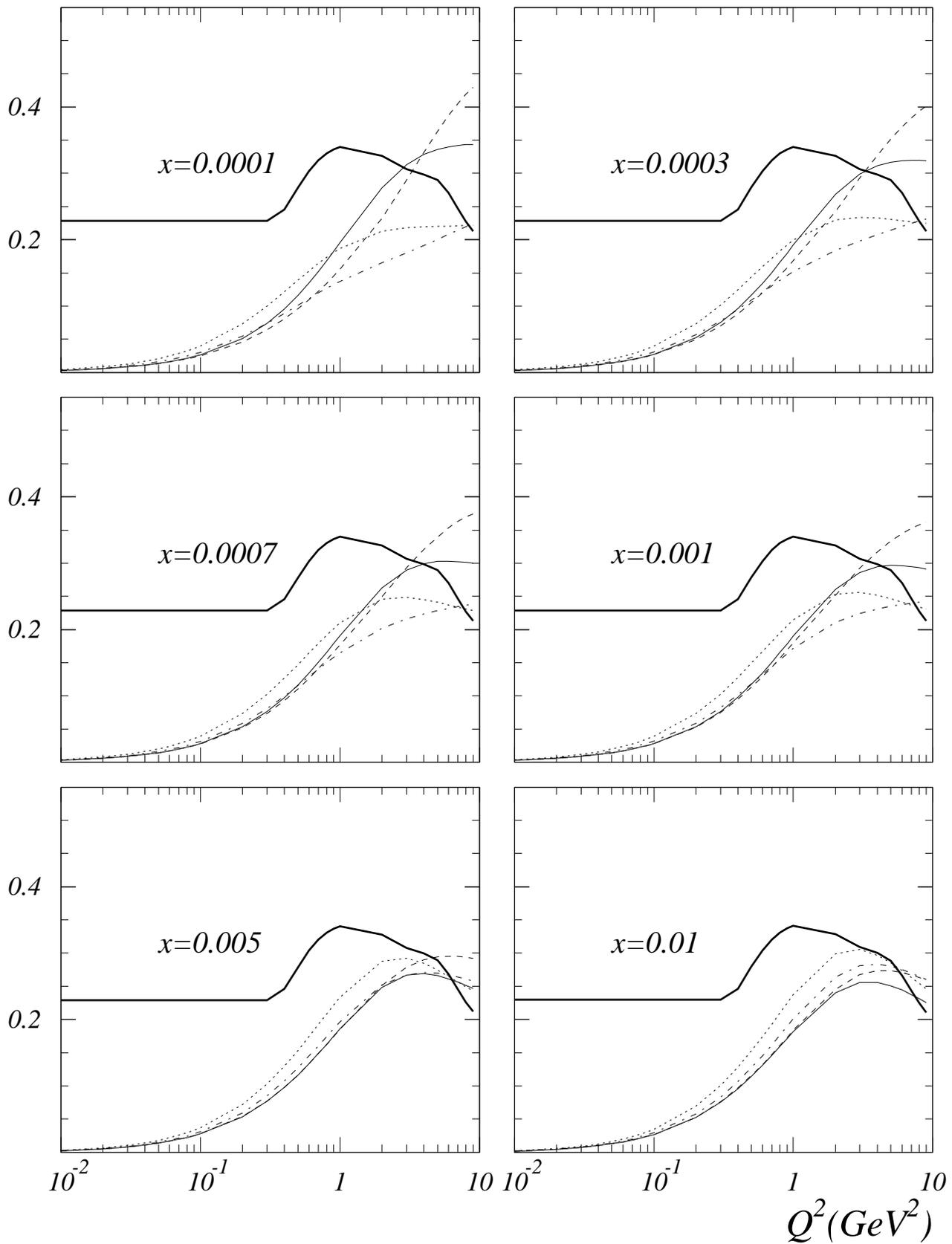
{\bf Fig.4} The ratio $R$ plotted as a function of $Q^2$ for different
values of $x$ with the higher twist effect included and compared with
the SLAC parametrisation (thick solid lines).
Solid and dashed lines correspond to the off--shell and on--shell
approximation with the GRV parametrisation of the parton distributions \cite{GRV}.
The dotted and dashed-dotted lines correspond to off--shell and
on--shell approximation calculated using MRS(A) partons \cite{MRSA}.


\begin{thebibliography}{999}
\bibitem{LANDSHOFF}P.V. Landshoff, J.C. Polkinghorne, K.D. Short,
Nucl. Phys. {\bf B28} (1971), 225.
\bibitem{BADELEK3} M. Arneodo, {\it{et al.}} (NMC), preprint CERN-PPE-95-138,
(1995).
\bibitem{BADELEK4} D. Adams, {\it{et al.}} (SMC), Phys. Lett. B357 (1995) 248.
\bibitem{AKMS} A.J. Askew, J. Kwieci\'nski, A.D. Martin, P.J. Sutton,
Phys. Rev. {\bf D47} (1993) 3775; Phys. Rev. {\bf D49} (1994) 4402.
\bibitem{BBJKAS} B. Bade\l{}ek, J. Kwieci\'nski, A. Sta\'sto, DTP/96/16.
\bibitem{KT} S. Catani, M. Ciafaloni and F. Hautmann, Phys. Lett. 
{\bf B242} (1990) 97; Nucl. Phys. {\bf B366} (1991) 657; 
J.C. Collins and R.K. Ellis,\ Nucl. Phys. {\bf B 360} (1991) 3; 
S. Catani and F. Hautmann, Nucl. Phys. {\bf B427} (1994) 475;
M. Ciafaloni, Phys. Lett. {\bf 356} (1995) 74.
\bibitem{KTBL} J. Bl\"umlein,  J. Phys. {\bf G19} (1993) 1623.
\bibitem{BFKL} E.A. Kuraev, L.N.Lipatov and V.S. Fadin, Zh. Eksp. Teor. Fiz. 
{\bf 72} (1977) 373 ( Sov. Phys. JETP {\bf 45} (1977) 199); 
Ya. Ya. Balitzkij and L.N. Lipatov, Yad. Fiz. {\bf 28} (1978) 1597
( Sov. J. Nucl. Phys. {\bf 28} (1978) 822);  
J.B. Bronzan and R.L. Sugar, Phys. Rev. {\bf D17} (1978) 585; 
T. Jaroszewicz, Acta. Phys. Polon. {\bf B11} 
(1980) 965. 
\bibitem{LIP} L.N. Lipatov, in " Perturbative QCD " edited 
by A.H. Mueller, (World Scientific, Singapore, 1989), p. 441.
\bibitem{AP} G. Altarelli, G. Parisi, Nucl. Phys. {\bf B126} (1977) 298.
\bibitem{KTJKAM} J. Kwieci\'nski, A.D. Martin,
Phys. Lett. {\bf B353} (1995) 123.
\bibitem{RGR}R.G. Roberts,  "Structure of Proton" (1990)  Cambridge
University Press.
\bibitem{AM}G. Altarelli, G. Martinelli,  Phys. Lett. {\bf B76} (1978) 89.
\bibitem{BODEK} A. Bodek, Proceedings of the Rencontre de Blois, 
"The Heart of Matter", Blois, France, June 25-25 1994, edited by J.F. Mathiot 
and J. Tr\^an Thanh V\^an, Editions Fronti\`eres, 1995.
\bibitem{SLACR} L.W.Whitlow et al., Phys. Lett., {\bf B250} (1990) 193;
\bibitem{GRV} M. Gl\"uck, E. Reya and A. Vogt, Z. Phys.  {\bf C67}
(1995) 433.
\bibitem{MRSA} A.D. Martin, R.G. Roberts and W.J. Stirling ,
 Phys. Rev.{\bf D50} (1994) 6734;  Phys. Lett. 354 (1995) 155.
\bibitem{BLUEMLEIN} J. Bl\"umlein, DESY 94-149, hep-ph/9408377.
\bibitem{BBJK} B. Bade\l{}ek, J. Kwieci\'nski,  Phys. Lett.
{\bf B295} (1992) 263.
\end{thebibliography}
\end{document}